\documentstyle[aps,preprint]{revtex}

\begin{document}
\textheight 240truemm
\draft
\tightenlines
\title
{The investigation of the high frequency hopping conductivity
in two- and three-dimensional electron gas by an acoustic method.}

\author{I.L.Drichko, A.M.Diakonov,
I.Yu.Smirnov, A.V.Suslov}
\address{A.F.Ioffe Physicotechnical Institute of RAS, 194021, St.Petersburg, Russia.}

\maketitle

\begin{abstract}
High-frequency (HF) conductivity ($\sigma _{hf}$) 
measured by an acoustical method has been studied in GaAs/AlGaAs
heterostructures in a linear and nonlinear regime on acoustic power. It has
been shown that in the quantum Hall regime at magnetic fields corresponding
to the middle of the Hall plateaus the HF conductivity is determined by the
sum of the conductivity of 2-dimensional electrons in the high-mobility
channel and the hopping conductivity of the electrons in the doped thick
AlGaAs layer. The dependence of these conductivities on a temperature is
analyzed. The width of the Landau level broadened by the impurity random
potential is determined.
\end{abstract}


\section{Introduction} 
If one places a semiconducting
heterostructure over a piezoelectric platelet along which acoustic surface
wave (SAW) is being propagated, the SAW undergoes additional attenuation
associated with the interaction of the electrons of heterostructure with the
electric field of SAW. This is the basis of the acoustic method pioneered by
Wixforth \cite{1} for the investigation of GaAs/AlGaAs heterostructures. In
contrast to \cite{1} in the present work it has been found that in a
GaAs/AlGaAs heterostructure ($\mu =1.28\cdot 10^5cm/Vs$, $n=6.7\cdot
10^{11}cm^{-2}$) in the quantum Hall regime the conductivities $\sigma _{hf}$
measured by an acoustic method do not coincide with $\sigma _{dc}$ obtained
from the direct-current measurements: $\sigma _{dc}$ =0 , whereas $\sigma
_{hf}$ has a finite value depending on a temperature, magnetic field and SAW
intensity. And this is the main objective of the present work to elucidate
the nature of this HF conductivity.

The theoretical absorption coefficient can be presented in the following way
\cite{2}:
\begin{eqnarray}
\Gamma=8.68 \frac{K^2}{2}
kA\frac{(4\pi \sigma_{xx}/\epsilon_s v)c(k)}
{1+[(4\pi \sigma_{xx}/\epsilon_s v)c(k)]^2},
A=8b(k)(\epsilon_1+\epsilon_0)\epsilon_0^2\epsilon_sexp(-2ka),\label{eq:gam}
\end{eqnarray}
where $K^2$ is the electromechanical coupling coefficient of piezoelectric
substrate, $k$ and $v$ are the SAW wavevector and the velocity respectively,
$a$ is the vacuum gap width between the lithium niobate platelet and the
sample, $\sigma _{xx}$ is the dissipative HF conductivity of 2DEG,
$\epsilon _1$,
$\epsilon _0$ and $\epsilon _s$ are the dielectric constants of lithium
niobate, vacuum and semiconductor respectively; $b$ and $c$ are some complex
functions of $a$, $k$, $\epsilon _1$, $\epsilon _0$ and $\epsilon _s$. When $%
(4\pi \sigma _{xx}/\epsilon _sv)c(k)$=1, $\Gamma $ achieves the maximum $%
\Gamma _m$. In the case when $\sigma _{xx}/v<<$ 1 (quantum Hall regime) $%
\Gamma \propto \sigma_{xx}$. Thus, the SAW electronic attenuation may be taken
as a measure of the heterostructure conductivity.

The measurements were carried out in a temperature range 1.5-4.2K , and
magnetic fields $B$ up to 7T, with an acoustic frequency in the range of
30-150 MHz. Two kinds of measurements were performed: for the first kind the
acoustic power maintained low enough to provide the linearity of results,
for the second kind a nonlinear behavior on acoustic power $P$ level was
studied intensively at 1.5K.

\section{Experimental results and discussion}
Fig.1 illustrates the
experimental dependences of $\Gamma $ on $B$. As long as the SAW attenuation
factor is determined by the sample conductivity, quantizing of the electron
spectrum in a magnetic field, leading to the SdH oscillations should result
in similar peculiarities of the curves of Fig 1. In the present work the
experimental data for the magnetic fields corresponding to the
quantum Hall regime will be analyzed.

Fig.2 presents the $\Gamma/\Gamma_m(T)$ dependences (f=30 MHz) at magnetic
fields 4.8, 3.6, and 2.9T corresponding to the attenuation minima (or the
middle of the Hall plateau), which are deduced from the
curves of Fig.1, measured at different $T$ and $f$. As one can see from the
figure, as $T$ grows, in a certain temperature range $\Gamma$ does not
depend on a temperature, but at higher temperature begins to grow
exponentially, the stronger $B$, the higher $T$ at which the growth starts.

Such a dependence could not be explained, if one supposes that $\Gamma (T)$
is determined solely by the 2-dimensional electrons. Indeed, in the quantum
Hall regime when the Fermi level is in the middle between two Landau levels,
in the temperature range 1.5-4.2 K the temperature dependence of the
2-dimensional conductivity $\sigma _2$ (and corresponding attenuation
$\Gamma _2$) is governed by the activation of electrons from the bound states
at the Fermi level to the upper Landau band, i.e. $\Gamma _2\propto
\sigma_2\propto exp(-\Delta E_g/kT)$.

The dependence of $\Gamma$ on temperature could be explained, if one
supposes that at these levels of $B$ the attenuation adds up both from the
SAW attenuation $\Gamma _2$ by the 2-dimensional electrons, and $\Gamma _{h}
$, due to the electrons, localized on the impurities in the
quasi-3-dimensional AlGaAs layer, which supplies carriers to the 2-dimensional
channel.

According to \cite{Pollak,4,5} in a transverse magnetic field $\Gamma
_{h}\propto\sigma_{h}\propto \omega B^{-2}$ and does not depend on a
temperature. Thus the independence of $\Gamma$ on a temperature at low $T$
could be interpreted as a dominance of the conductivity of the
quasi-3-dimensional layer. As $T$ rises the 2-dimensional conductivity
becomes prevailing. Based on the above arguments, $\Gamma _2$ has been
determined as a difference between the experimentally obtained values of $%
\Gamma$ and $\Gamma_{h}$. $\Gamma _{h}$ is equal to the $\Gamma$ in the
temperature range where $\Gamma$ is temperature-independent at the magnetic
fields 4.8 and 3.6T ($\Gamma _2<<\Gamma _h$). At $B$=2.9T, when there is no
temperature flattening , $\Gamma _{h}$ was obtained, using the assumption
that $\Gamma _h\propto 1/B^2$, \cite{5}. From the plotted dependence $\lg
\Gamma _2 (1/T)$ the activation energy $\Delta E_g$ has been found at $B$=
4.8, 3.6, 2.9 T. Fig.3 shows the $\Delta E_g$ dependence on $B$. The
activation energy obtained from the curves at different magnetic fields
appeared to be less than the corresponding energy $\hbar\omega _c/2$. This
may be due to  the Landau level broadening {\it A} caused by the random
fluctuation potential.

Supposing $\Delta E_g=\hbar \omega _c/2-A/2$, where $\hbar $ is the Plank
constant, $\omega _c=eB/m^{*}c$ is the cyclotron frequency, from the ordinate
cut-off point of the $\Delta E_g(B)=0$ curve for $B$=2.1T one could obtain
$A$, which appeared to be 3.4 meV. The slope of the $\Delta E_g(B)$ line in
fig.3 appeared to be 0.72 1/$B$, which by 10\% differs from the value
$e/m^{*}c$=0.8 1/$B$,
if $m^{*}=0.07m_0$ for GaAs ($m_0$ - free electron mass).

The dependence of $\Gamma /\Gamma _m$ on $P$ at the same magnetic field is
shown in fig.4. It can be seen from the figure, that $\Gamma $ increases
with the increase of $P$. Keeping in mind that at given magnetic fields
$\Gamma _2$ is determined by the electron activation to the upper Landau
band from the Fermi level, one could suppose that Frenkel-Pool effect i.e
the activation energy decrease in electric field $E$ of a SAW \cite{Stilb}
is operative in this case.
\begin{eqnarray}
\Gamma_2\propto\sigma_2\propto n(T,E)=n_0exp(2e^{3/2}E^{1/2}
\epsilon_s^{-1/2}/kT),
\label{eqn2}
\end{eqnarray}
where $n_0$ -carrier density in the upper Landau level at a linear approach
at 1.5K. $Ln\Gamma _2$ plotted against $E^{1/2}$, where $\Gamma _2=\Gamma
-\Gamma _{h}$, and $E$ is the electric field of SAW \cite{6}, which can be
presented by a straight line, confirms our model.

\section{Conclusion}
High-frequency conductivity $\sigma _{hf}$ of a
GaAs/AlGaAs heterostructure with a moderate mobility has been studied in the
quantum Hall regime ($\sigma _{dc}=0)$ by an acoustic contactless method. It
has been shown that up to T=2K $\sigma _{hf}$ is determined by the hopping
conductivity in thick, quasi three-dimensional layers of AlGaAs well
described by a Pollak-Geballe model \cite{Pollak}. The two-dimensional HF
conductivity in the quantum Hall regime is governed by the electron
activation into the upper Landau band. The small value of this
conductivity allows one to suppose it also to be of a hopping nature, but
this point needs further arguments to be proved.

The work was supported by the RFFI N95-02-04066-a and MINNAUKI N97-1043
grants.

\begin{figure}
\caption{The experimental dependence of absorption coefficient 
$\Gamma$ on $B$ at $f=30MHz$
at $T=4.2K$.}
\end{figure}

\begin{figure}
\caption{The dependences of $\Gamma/\Gamma_m$ ($f=30MHz$) on $T$ at
magnetic fields: $1$-4.8T, $2$-3.6T, $3$-2.9T.}
\end{figure}

\begin{figure}
\caption{The dependences of $\Delta$ on magnetic field for two
frequencies: $1$-30MHz, $2$-150MHz.}
\end{figure}

\begin{figure}
\caption{The dependences of $\Gamma /\Gamma _m$ on the RF-generator
output power in magnetic fields: $1$-4.8T, $2$-3.6T, $3$-2.9T.}
\end{figure}

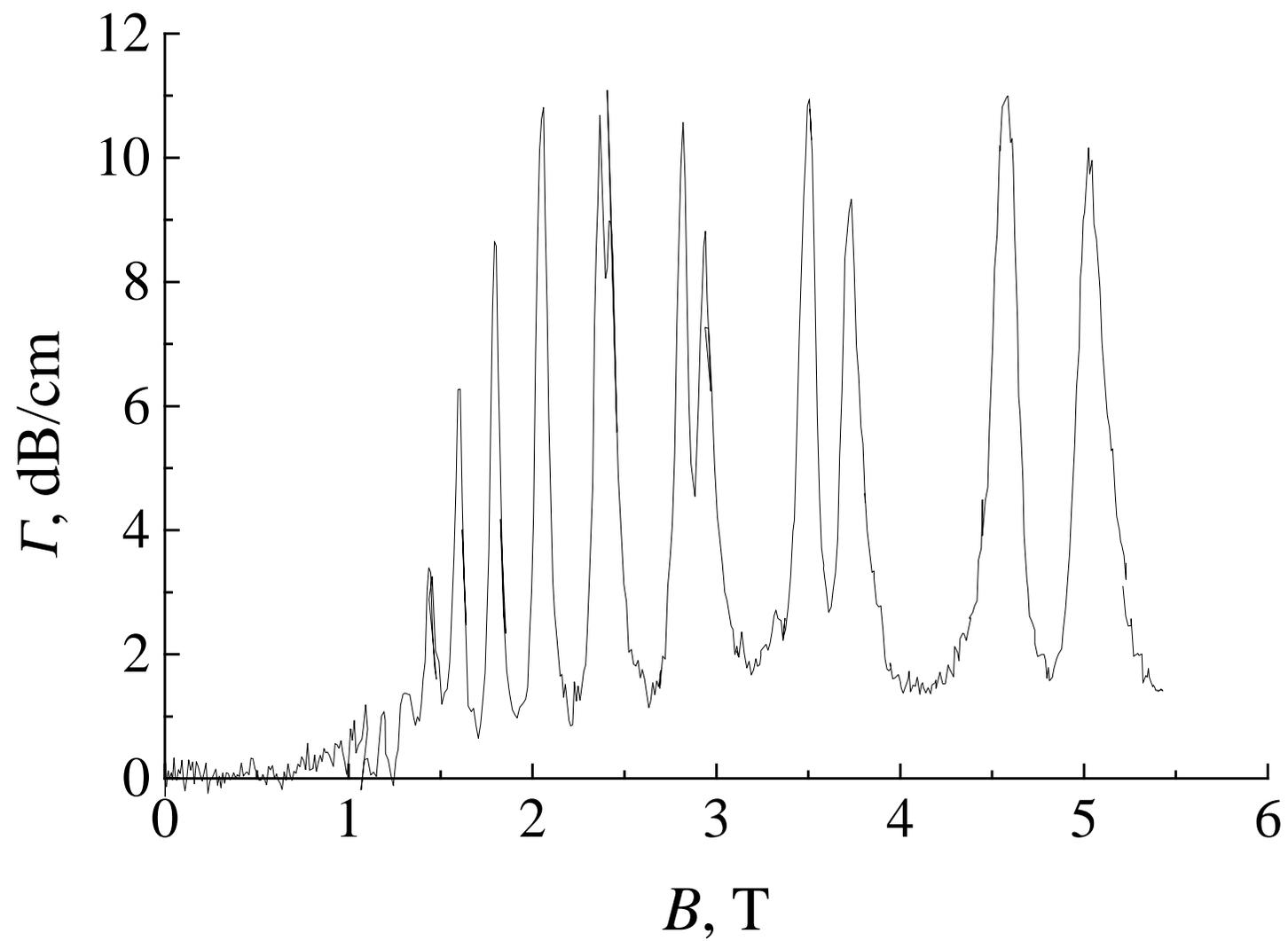
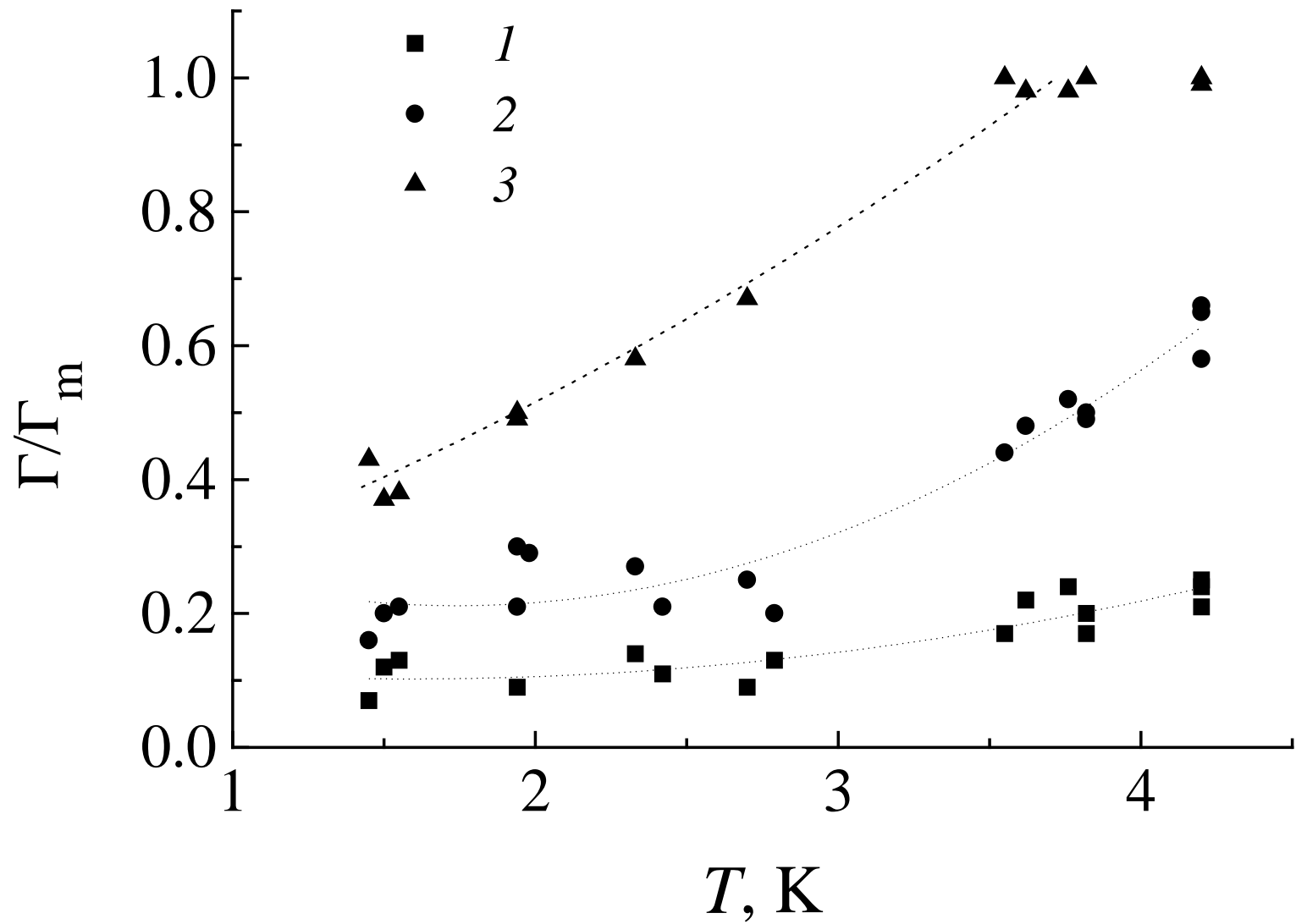
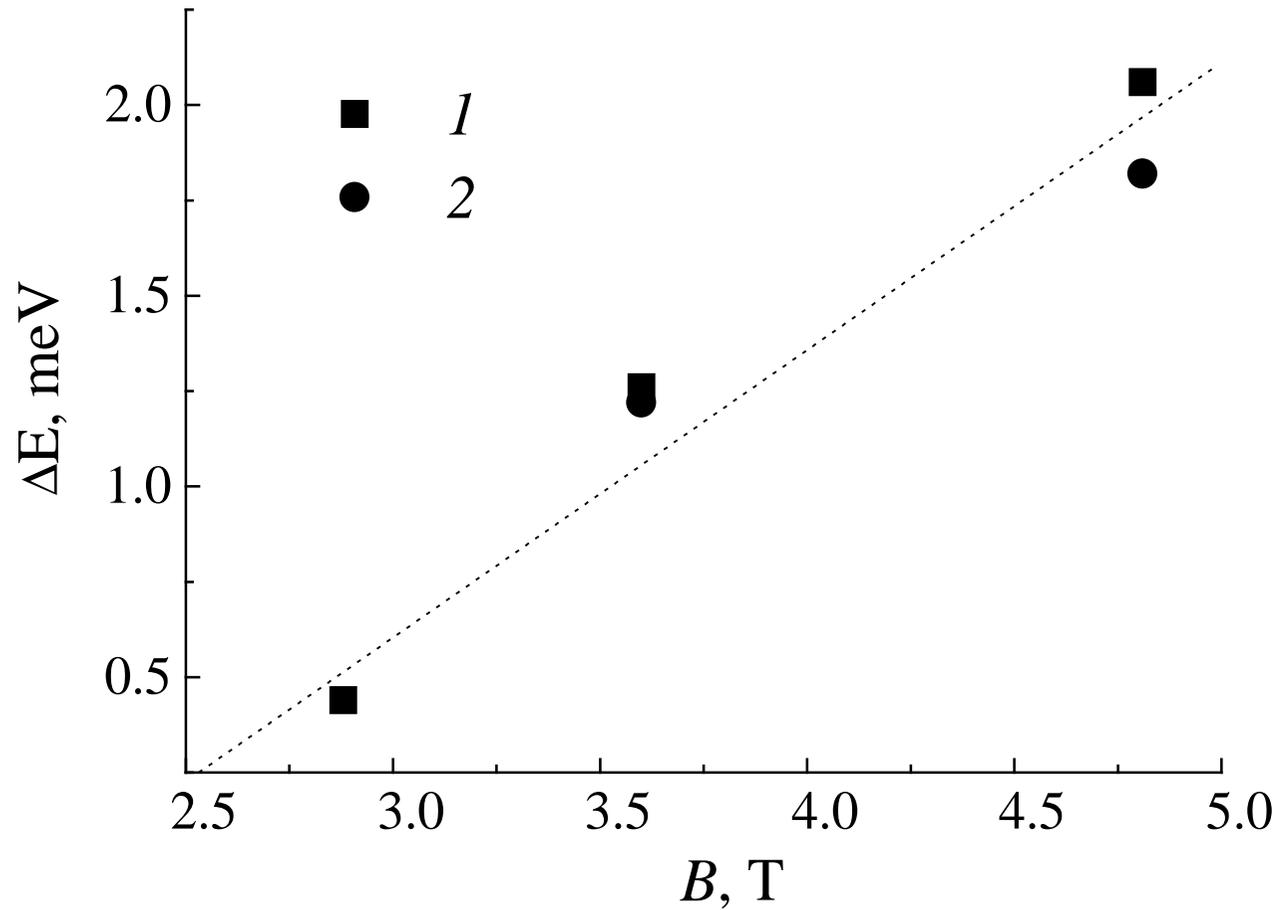
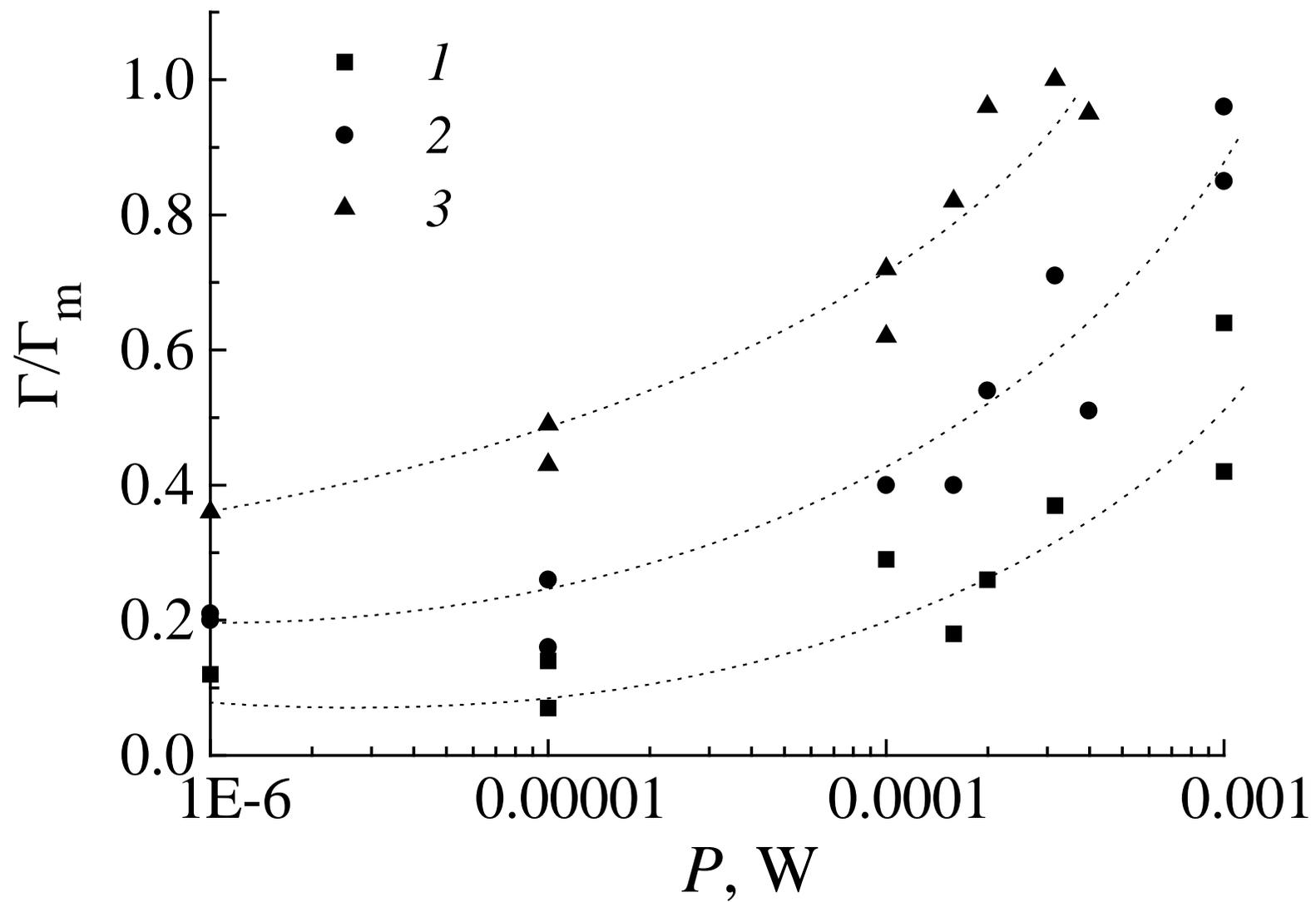
\end{document}